\documentclass[a4paper,fleqn,usenatbib]{mnras}

\usepackage{newtxtext,newtxmath}
\usepackage[T1]{fontenc}
\usepackage{ae,aecompl}

\usepackage{graphicx}	% Including figure files
\usepackage{amsmath}	% Advanced maths commands
\usepackage{amssymb}	% Extra maths symbols

\title[\hbox{[OI]}63\,$\mu$m in main-sequence galaxies]{Observations of
  \hbox{[OI]}63\,$\mu$m  line emission in main-sequence galaxies at $z \sim 1.5$\thanks{{\it Herschel} is an
  ESA space observatory with science instruments provided by
  European-led Principal Investigator consortia and with important
  participation from NASA.} }

\author[J. Wagg et al.]{
\parbox[t]{\textwidth}{
J. Wagg,$^{1}$\thanks{E-mail: j.wagg@skatelescope.org}
M. Aravena,$^{2}$ D. Brisbin,$^{3}$ I. Valtchanov,$^{4}$ C. Carilli,$^{5}$
E. Daddi,$^{6}$ H. Dannerbauer$^{7,8}$ R. Decarli,$^{9}$
T. Diaz-Santos,$^{2,10}$ D. Riechers,$^{11,12}$ M. Sargent$^{13}$ and 
F. Walter$^{12}$}\\\\
$^{1}$SKA Organization, Lower Withington Macclesfield, Cheshire SK11 9DL, UK\\
$^{2}$N\'ucleo de Astronom\'ia, Facultad de Ingenier\'ia y Ciencias,
Universidad Diego Portales, Av. Ej\'ercito 441, Santiago, Chile\\
$^{3}$Atacama Large Millimeter/Submillimeter Array, Joint ALMA
Observatory, Alonso de C\'ordova 3107, Vitacura 763-0355, Santiago,
Chile\\
$^{4}$Telespazio Vega UK for ESA, European Space Astronomy Centre, Operations Department, 28691 Villanueva de la Ca\~nada, Spain \\
$^{5}$National Radio Astronomy Observatory, P.O. Box 0, Socorro, NM 87801, USA \\
$^{6}$CEA, IRFU, DAp, AIM, Universit\'e Paris-Saclay, Universit\'e
Paris Diderot,  Sorbonne Paris Cit\'e, CNRS, F-91191 Gif-sur-Yvette,
France\\
$^{7}$Instituto de Astrof\'isica de Canarias (IAC), 38205, La Laguna,
Tenerife, Spain \\
$^{8}$Universidad de La Laguna, Dpto. Astrof\'isica, 38205, La Laguna, Tenerife, Spain  \\
$^{9}$INAF - Osservatorio di Astrofisica e Scienza dello Spazio, Via Gobetti 93/3, 40129, Bologna, Italy  \\
$^{10}$Chinese Academy of Sciences South America Center for Astronomy
(CASSACA), National Astronomical Observatories, CAS, Beijing 100101,
People's Republic of China \\
$^{11}$Department of Astronomy, Cornell University, 220 Space Sciences Building, Ithaca, NY, 14853, USA \\
$^{12}$Max Planck Institute for Astronomy, K\"onigstuhl 17, D-69117
Heidelberg, Germany  \\
$^{13}$Astronomy Centre, Department of Physics and Astronomy, University of Sussex, Brighton BN1 9QH, UK \\
}

\date{Accepted XXX. Received YYY; in original form ZZZ}

\pubyear{2020}

\begin{document}
\label{firstpage}
\pagerange{\pageref{firstpage}--\pageref{lastpage}}
\maketitle

\begin{abstract}
We present \textit{Herschel}-PACS spectroscopy of four main-sequence
star-forming galaxies at $z \sim 1.5$. We detect \hbox{[OI]}63\,$\mu$m
line emission in BzK-21000 at $z=1.5213$, and measure a line
luminosity, L$_{\rm [OI]63\,\mu m} = (3.9\pm 0.7)\times
10^9$~L$_{\odot}$. Our PDR modelling of
the interstellar medium in
BzK-21000 suggests a UV radiation field strength, $G\sim 320G_0$, and
gas density, $n \sim 1800~$cm$^{-3}$, consistent with previous LVG
modelling of  the molecular CO line excitation.     
The other three targets in our sample are individually
undetected in these data, and we perform a spectral stacking analysis which
yields a detection of their average emission and an \hbox{[OI]}63\,$\mu$m line
luminosity, L$_{\rm [OI]63\,\mu m} =
(1.1\pm 0.2)\times10^9$~L$_{\odot}$. 
We find that the implied luminosity ratio,
L$_{\rm [OI]63\,\mu m}/$L$_{\rm IR}$, of the undetected BzK-selected
star-forming galaxies broadly agrees with that of low-redshift
star-forming galaxies, while BzK-21000 has a similar ratio to that of a dusty
star-forming galaxy at $z \sim 6$. The high \hbox{[OI]}63\,$\mu$m line
luminosities observed in BzK-21000 and the $z \sim 1 -3$ dusty and submm luminous star-forming
galaxies may be associated with extended reservoirs of low density, cool neutral gas. 
\end{abstract}

\begin{keywords}
galaxies: high-redshift -- galaxies: star formation -- galaxies: ISM -- infrared: ISM
\end{keywords}

%%%%%%%%%%%%%%%%%%%%%%%%%%%%%%%%%%%%%%%%%%%%%%%%%%

%%%%%%%%%%%%%%%%% BODY OF PAPER %%%%%%%%%%%%%%%%%%

\section{Introduction}

The Cosmic star-formation rate density is known to have been
significantly higher in the past (see review by Madau \& Dickson 2014). This galaxy formation
will have been fed by larger molecular gas masses than what is observed in
present day galaxies, as
confirmed by observations of redshifted CO line emission (see review by Carilli
\& Walter 2013). A significant
contributor to the star-formation rate density of the Universe at the peak epoch of galaxy build up ($z=1-3$) was the population of massive star-forming main-sequence galaxies
(Brinchmann et al.\ 2004; Daddi et al.\ 2007; Elbaz et al.\ 2007;
Rodighiero et al.\ 2011; 
Sargent et al.\ 2012). Several
studies of molecular CO line emission have concluded that these galaxies have long depletion timescales,
high molecular gas fractions and typically evolve secularly with
redshift (e.g. Daddi et al. 2010a,b; Tacconi et al.\ 2010, 2013, 2018;
Freundlich et al.\ 2019).  In general, 
considerable advances have been made in our understanding of the molecular and atomic interstellar
medium (ISM) properties of main-sequence galaxies at $z >  1$
(Dannerbauer et al.\ 2009;
Aravena et al.\ 2010, 2019; Tacconi et al.\ 2013, 2018; Daddi et al.\ 2015;
Genzel et al.\ 2015; Decarli et al.\ 2016; Valentino et al.\ 2018,
2020; Zanella et al.\ 2018; Brisbin et al.\ 2019).

Owing to their brightness at rest-frame FIR wavelengths,  the ionized
and neutral species of Carbon, Nitrogen and Oxygen are powerful diagnostic lines for
tracing the ISM of nearby and distant galaxies. When combined with
photo-dissociation region (PDR) models (Tielens \& Hollenbach 1985), 
measurement of the emission
from different lines provides a means to 
constrain quantities such as the ionization rate and metallicity of
the ISM of galaxies. Multi-level FIR transition lines have now been widely surveyed
in local galaxies, originally by \textit{ISO} (e.g. Luhman et al.\ 1998;
Malhotra et al.\ 2001) and more
recently using the PACS spectrometer (Poglitsch et al.\ 2010) on the
ESA \textit{Herschel Space Observatory}
(Pilbratt et al.\ 2010). The \hbox{[CII]}158\,$\mu$m line is typically the
brightest in star-forming galaxies, arising from ionized, and even neutral gas where
it is the main coolant (Wolfire et al.\ 2003). Another commonly
observed FIR line tracer of the ionized ISM is \hbox{[NII]}122\,$\mu$m. It has
the advantage that it can be found associated with lower excitation
gas, close to that observed in our own Galaxy (e.g. Goldsmith et al.\
2015; Herrera-Camus et al.\ 2016). Another major coolant of the ISM is \hbox{[OI]}63\,$\mu$m
(Wolfire et al.\ 2003). Owing to its high excitation temperature and
critical density, it can dominate the cooling in regions of starburst
activity (Kaufman et al.\ 1999, 2006; Brauher et al.\ 2008; Croxall
et al.\ 2012; Narayanan \& Krumholz 2017).
 When combined with measurements of the
\hbox{[CII]}158\,$\mu$m line intensity and FIR luminosity, the
\hbox{[OI]}63\,$\mu$m line intensity can constrain the FUV field, $G$, and the
gas density using PDR models. The luminosity in these FIR lines generally exhibits a deficit
in the most FIR luminous galaxies compared to the trend expected from
lower luminosity galaxies (e.g. Malhotra et al.\ 2001;
Graci{\'a}-Carpio et al.\ 2011; Diaz-Santos et al.\ 2017). This has made the emission from lines like
\hbox{[OI]}63\,$\mu$m more challenging to detect at high-redshifts.

Studies of the \hbox{[OI]}63\,$\mu$m line in the distant Universe have been further
limited by the opacity of the atmosphere. Space-based observations provide the most promising route
to detecting this line in either emission or absorption during the $z \sim 1 - 2$ epoch of peak
star-formation.  The
\textit{Herschel}-PACS spectrometer enabled observations of the
\hbox{[OI]}63\,$\mu$m line emission in high-redshift, 
submm-selected starburst galaxies (Sturm et al.\
2010; Coppin et al.\ 2012; Brisbin et al.\ 2015; Wardlow et
al.\ 2017; Zhang et al.\ 2018), which confirm the FIR line
deficit observed in nearby luminous and ultraluminous galaxies. Most
recently, Rybak et al.\ (2020) have made a ground-based detection of
\hbox{[OI]}63\,$\mu$m in G09.83808, a dusty z$\sim$6 galaxy, and from 
these data they infer a gas density, $n = 10^4$~cm$^{-3}$, and FUV field
strength, G$= 10^4$~G$_0$\footnote{Note that $G_0$ is the Habing
    field unit and is equal to $1.6 \times
    10^{-3}$~erg~s$^{-1}$~cm$^{-2}$.}. To date, few observations of this
line have been presented for lower redshift, main sequence
star-forming objects like the BzK galaxies.

Here, we present \textit{Herschel}-PACS spectroscopy of four
BzK-selected star-forming galaxies at $z \sim 1.5$. The paper is
organized as follows: in Section~2 we describe the sample selection
along with the \textit{Herschel}-PACS observations and data
analysis. In Section 3 we present our results and discussion,
including PDR modelling of the luminosity ratios. Finally, we conclude in Section 4. Throughout this work, we
adopt a flat $\Lambda$CDM cosmology with parameters measured by Planck Collaboration et al.\
(2016).

\section{Observations and Data Reduction}

\subsection{Selection of BzK galaxy sample}

Our targets were selected to be massive ($\rm log M_{stars}/M_{\odot} > 10.5$), disk galaxies at $z \sim 1.5$, detected in multiple CO
line transitions (Daddi et al.\ 2008,
2010a,b, 2015; Dannerbauer et al.\ 2009; Aravena et al.\ 2010). There
were four main-sequence galaxies with observations of multiple CO
line transitions at the time of the proposal, and all benefitted from a wealth
of multi-wavelength data covering UV-to-cm wavelengths (Capak et al.\
2004; Wirth et al.\ 2004; Barger et al.\ 2008; Magdis et al.\ 2010,
2012; Morrison et al.\ 2010; Teplitz et al.\ 2011). 
The data have been used to measure infrared
luminosities, L$_{\rm IR} \sim (1 -2)\times 10 ^{12}$~L$_{\odot}$, and
estimate star-formation rates of $SFR \sim 100 -
200$~M$_{\odot}$~yr$^{-1}$. With
the exception of BzK-17999 which has not been detected in
CO~\textit{J} =1-0 line emission, all of our sources have been
observed in CO~\textit{J} =1-0, 2-1, 3-2 and 5-4 line emission. 
Some of the observational properties of our targets are
provided in Table~\ref{tab:obs}.  

\begin{table*}
\begin{minipage}{1.0\textwidth}
%\scriptsize
\caption{Properties of the targets in our sample, along with the PACS
  observing times and resulting spectral measurements. The CO line
  redshifts are from Daddi et al.\ (2010a,b), while the 8-to-1000~$\mu$m infrared
  luminosities have been derived by fitting their infrared spectral energy
  distributions (Magdis et al.\ 2012).}\label{tab:obs}
\scriptsize
\hspace{-0.2in}
\begin{tabular}{lrrrlrcccc}
\hline
\multicolumn{1}{c}{Source name} & \multicolumn{4}{c}{\underline{Source
                                  properties}} &  \multicolumn{3}{c}{\underline{Observation
                                                 Parameters}} &
                                                                \multicolumn{2}{c}{\underline{
                                                               [OI]63\,$\mu$m Spectral Measurements}} \\
 & \multicolumn{1}{c}{R.A.} & \multicolumn{1}{c}{Dec.} &  
   \multicolumn{1}{c}{z$_\mathrm{CO}$}  & \multicolumn{1}{c}{L$_{IR}$} & \multicolumn{1}{c}{OD} & \multicolumn{1}{c}{ObsID} 
& \multicolumn{1}{c}{Obs. time} &
                      \multicolumn{1}{c}{Integrated intensity} & rms per 80~km~s$^{-1}$   \\
 & \multicolumn{2}{c}{(J2000)} & & $\times 10^{12}$~L$_{\odot}$  & &  & \multicolumn{1}{c}{(hrs)} &
    \multicolumn{1}{c}{(Jy~kms~s$^{-1}$)} & \multicolumn{1}{c}{(mJy)} \\
\hline
BzK-4171  &  12:36:26.516 & 62:08:35.25 & 1.4652 & 1.0 & 1140 & 1342247456 &  9.0  & $< 2.2$ (3-$\sigma$) & 3.5 \\ % 5.7 mJy
BzK-21000 &  12:37:10.597 & 62:22:34.60 & 1.5213 & 2.1 & 1132 & 1342247133 & 7.5 &  15.3$\pm$2.7 &  9.1\\
BzK-16000 &  12:36:30.120 & 62:14:28.00 & 1.5250 & 0.7 & 1306 & 1342256932,  1342247133  & 9.6  & $<4.8$ (3-$\sigma$) & 12.9 \\ %21.2 mJy
BzK-17999 &  12:37:51.819 & 62:15:20.16 & 1.4139 & 1.1 & 1133  & 1342247156  & 6.3  & $< 9.4$ (3-$\sigma$) & 16.5 \\ % 29.7 
\hline
\end{tabular}
\normalsize
\end{minipage}
\end{table*}

\subsection{Observing setup and pipeline processing}

Observations were made with the PACS integral field unit (IFU) spectrometer on board the
\textit{Herschel Space Observatory} (OT2\_maravena\_3, PI: M. Aravena)
 during June and December, 2012. At the redshifts of our targets,
the \hbox{[OI]}63\,$\mu$m line will be redshifted to the 103 to
190$\mu$m R1 band, which we have used along with the high spectral sample density mode. The sky
background subtraction was achieved using a chop-nod technique and the
total on-sky observing times are given in Table~\ref{tab:obs}. The 
PACS integral field spectrometer consisted of 5$\times$5 spatial
pixels, where each is connected to two arrays of 16 spectral pixels.
Each spectrometer spatial pixel (or spaxel) has an approximate size of $9\farcs 4 \times 9\farcs 4$ at these wavelengths,
and the line emission is unresolved over these angular
scales (Daddi et al.\ 2010a,b). 

Data reduction was performed using the PACS data reduction and
calibration pipeline. We follow a similar recipe to that of Coppin et
al.\ (2012) in our data processing, using \textit{Herschel}
Interactive Processing Environment (HIPE v12.1.0; Ott 2010) and 
calibration tree version 58. This pipeline scales the continuum in
each pixel to the median value in order to perform the flat-field correction, and then
subsequently combines the nods for sky removal. As our sources are
expected to be unresolved at the resolution of these observations,
the spectra are extracted at the central pixel position. The spectra
have a resolution of $\sim$40~km~s$^{-1}$ prior to
resampling. No continuum emission is detected from the targets
  in our sample.  

\begin{figure*}
\begin{minipage}[b]{0.45\linewidth}
\centering
\includegraphics[width=\textwidth]{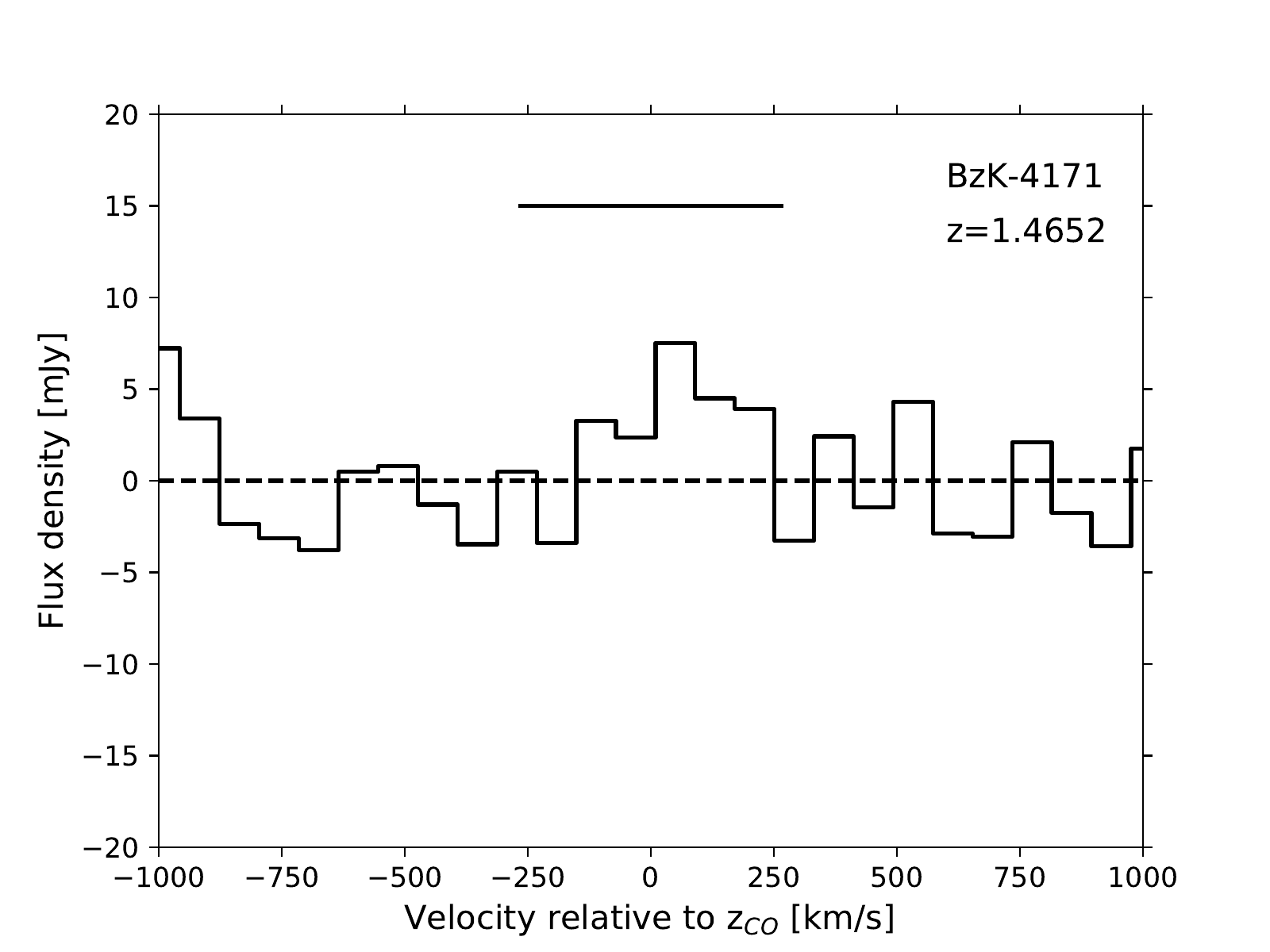}
\end{minipage} 
\begin{minipage}[b]{0.45\linewidth}
\centering
\includegraphics[width=\textwidth]{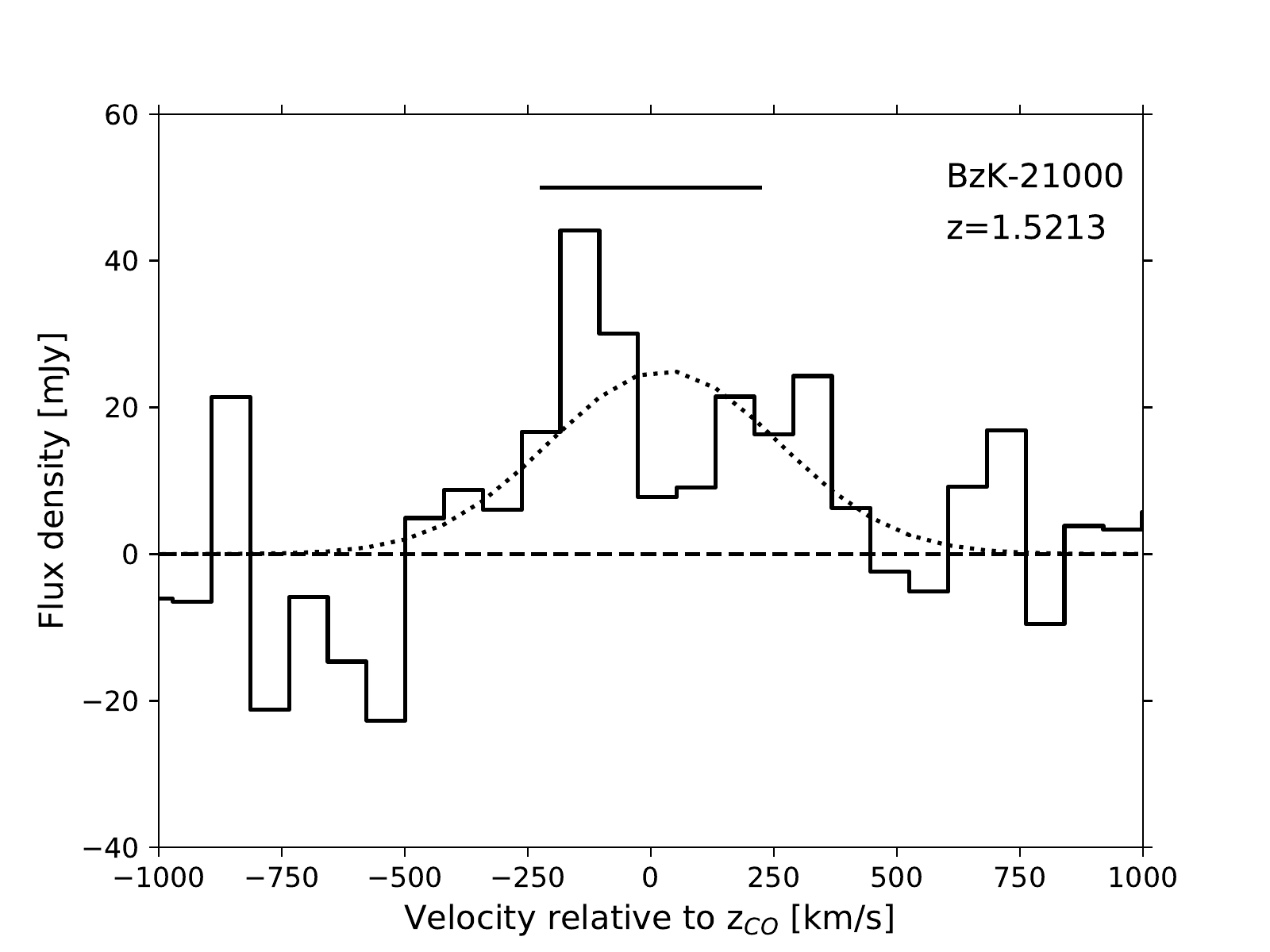}
\end{minipage} 
\begin{minipage}[b]{0.45\linewidth}
\centering
\includegraphics[width=\textwidth]{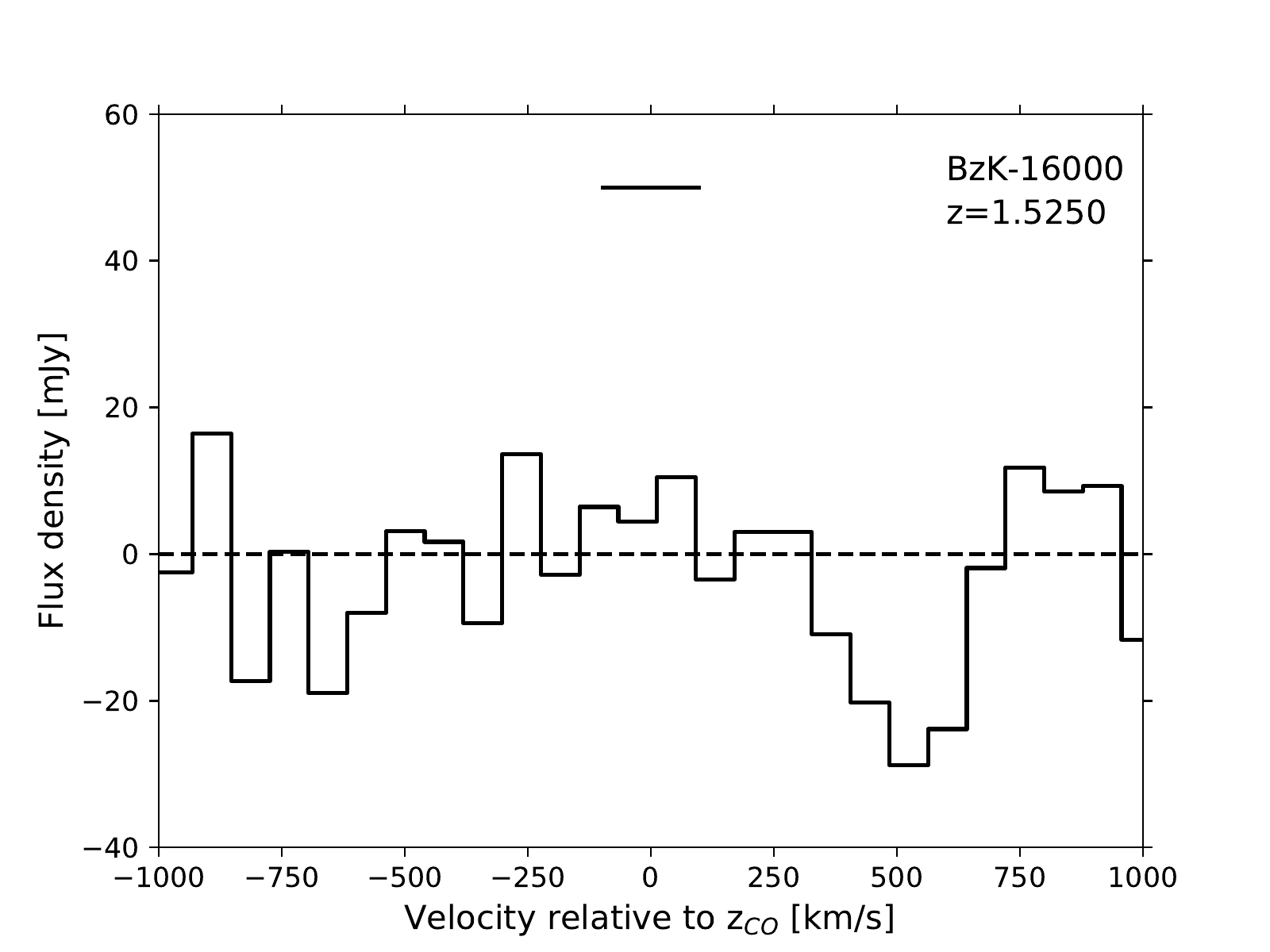}
\end{minipage} 
\begin{minipage}[b]{0.45\linewidth}
\centering
\includegraphics[width=\textwidth]{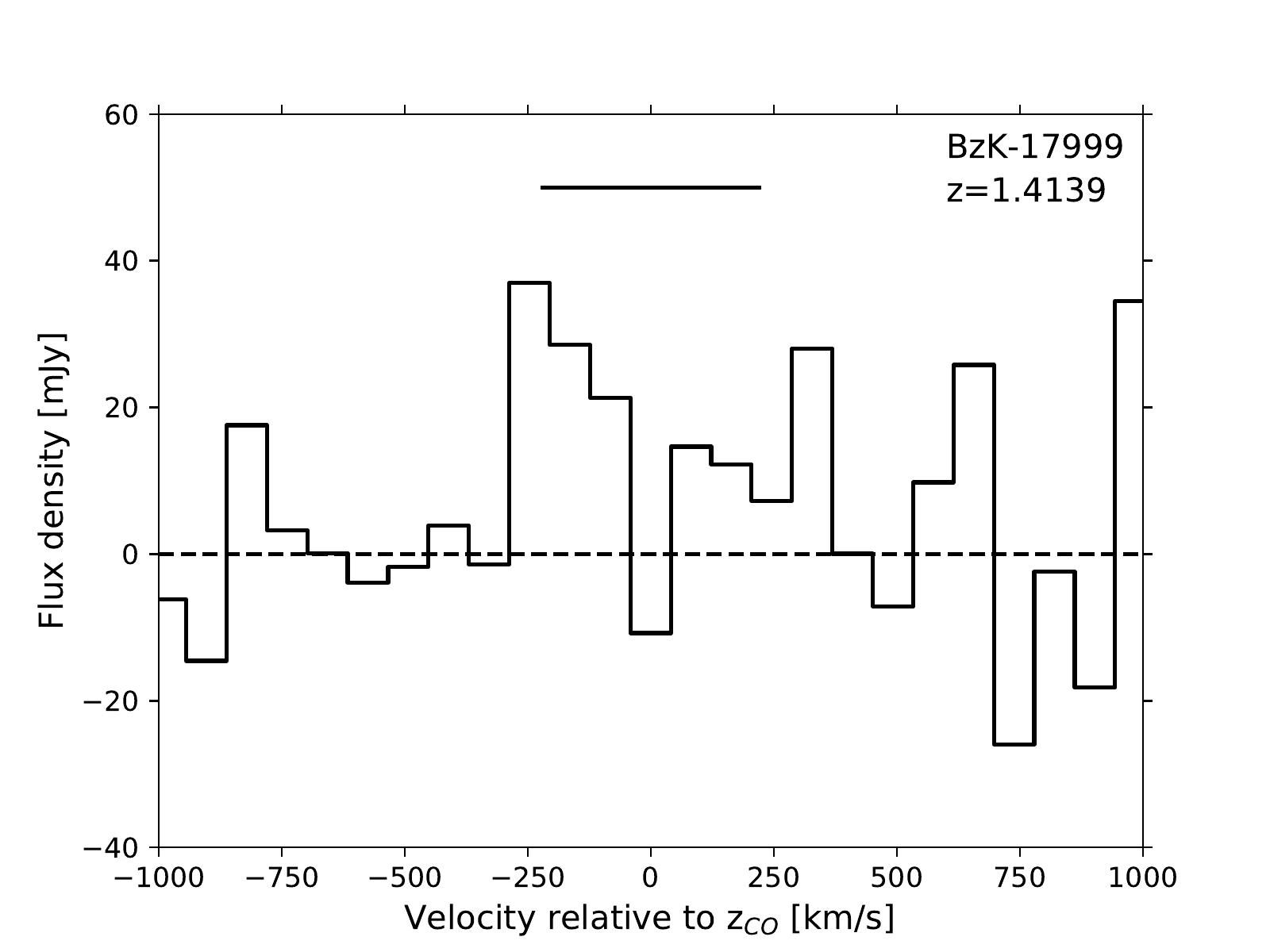}
\end{minipage} 
\caption{ \textit{Herschel}-PACS spectra of the four BzK-selected star-forming
  galaxies in our sample, tuned to the observed wavelength of the \hbox{[OI]}63\,$\mu$m
  line. Each spectrum is plotted relative to the redshift of the previously detected CO~\textit{J}=2-1 lines (Daddi et  al.\ 2010a), while the \textit{Solid horizontal} lines indicate the
  FWHM of the CO. The channels have been resampled and the spectral resolution corresponds to
  $\sim$80~km~s$^{-1}$. The only source which exhibits significant
  \hbox{[OI]}63\,$\mu$m line emission is BzK-21000 at z=1.5213, and
  the \textit{dotted line} shows a Gaussian fit with a peak of 25~mJy,
  FWHM$=550$~km~s$^{-1}$ and a velocity offset of 26~km~s$^{-1}$
  relative to the CO line.}
\label{fig:spectra}
\end{figure*}

The pipeline processed spectra are all modelled with a third order
polynomial fit to the regions of the data which are expected to be
line-free as indicated by the width of the previously detected CO
lines. This polynomial was deemed to be the lowest order
  which best represented the off-line spectral baseline structure
  while not introducing a spurious signal at the expected line wavelength. 
We also excluded channels at the outer edges of the
band in our fitting, as these are known to be noisy in PACS spectra (e.g. Coppin et
al.\ 2012). The total
region included in the line fitting corresponds to
$\sim$3000~km~s$^{-1}$. 
Figure~\ref{fig:spectra} shows plots of the spectra following
baseline subtraction, and resampling to $\sim$80~km~s$^{-1}$. 

\section{Results and Discussion }

\subsection{Detection of \hbox{[OI]}63\,$\mu$m in BzK-21000}

Of the four targets observed in our programme, we detect 
\hbox{[OI]}63\,$\mu$m  line emission only in BzK-21000 at $z =
1.5213$. The integrated line intensity is detected with a signal-to-noise
of 5.7-$\sigma$ (see Table~\ref{tab:obs}). Although there is a
hint of positive emission in the spectra of BzK-4171 and BzK-17999,
the significance of the integrated intensity is formally less than
3-$\sigma$ in each case. We calculate a line
luminosity for BzK-21000, L$_{\rm [OI]63\,\mu m} = (3.9\pm 0.7)\times
10^9$~L$_{\odot}$. When compared to galaxies in the nearby
  Universe, this line luminosity is only similar to that observed in the 
low-redshift ULIRG, NGC6240 (Diaz-Santos et al.\ 2017), a prototypical
dual AGN known to be
undergoing a merger (e.g. de Vaucouleurs et al.\ 1964; Komossa et al.\
2003; Wang et al.\ 2014). The line to infrared luminosity
ratio in NGC6240 is nearly 70\% higher than that observed in BzK-21000,
 possibly due to an absence of AGN activity in the latter. We
discuss this further below. 

The other three sources are individually undetected in these data and we
compute 3-$\sigma$ upper-limits to the integrated line intensities, $3
\sqrt{\Delta V_{\rm line} / \Delta V_{\rm chan}} \sigma_{\rm chan}
\Delta V_{\rm chan}$
(Isaak et al.\ 2004; Wagg et al.\ 2007). $\Delta V_{\rm chan}$ and
$\sigma_{\rm chan}$ are the channel linewidths and rms per channel,
respectively, while $\Delta V_{\rm line}$ is the assumed linewidth of the
\hbox{[OI]}63\,$\mu$m based on previous CO~\textit{J}=2-1 line
 measurements (Daddi et al.\ 2010a,b). We assume FWHM linewidths of 530, 194,
 and 440~km~s$^{-1}$ when calculating the \hbox{[OI]}63\,$\mu$m line intensity
 limits for BzK-4171, BzK-16000 and BzK-17999, respectively. Table~\ref{tab:obs}
 provides the calculated 3-$\sigma$ upper-limits on the integrated
 line intensity. 
 
\subsection{Spectral line stacking}

Although the \hbox{[OI]}63\,$\mu$m line emission is not detected
 in three of our targets, we perform a stacking analysis
of the spectra of the three undetected sources in 
order to determine if the line might be detectable
with more sensitive observations. Our approach is to
calculate the weighted mean of individual spectra, $S_i$, after first
 normalizing such that each is divided by the source FIR luminosity and
then scaled by 10$^{12}$~[L$_{\odot}$].

\begin{equation}
S_{stacked} = \frac{\sum_{i=1}^{n} w_i S_i}{\sum_{i=1}^{n} w_i}   
\end{equation}

\begin{figure}
\includegraphics[width=\columnwidth]{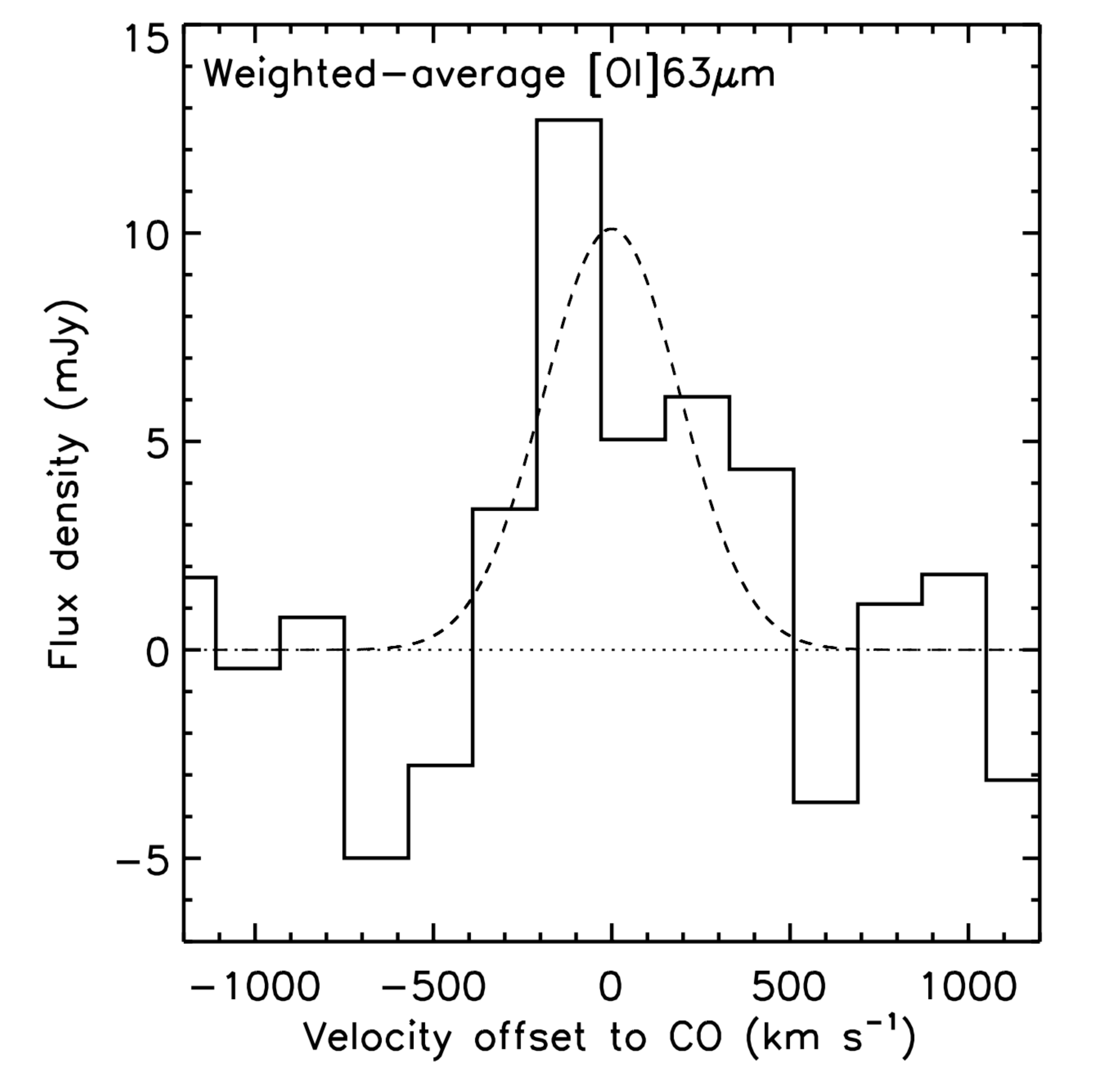}
   \caption{The \textit{solid line} shows the 
    stacked \hbox{[OI]}63\,$\mu$m spectrum for the three
     BzK-selected galaxies in our sample which are individually
     undetected in the data. A weighted average of the spectra is
     obtained by weighting the average by $1/\sigma_i$, where
     $\sigma_i$ is the rms calculated for each spectrum. The stacked
     \hbox{[OI]}63\,$\mu$m line emission is detected at 5.5-$\sigma$
     significance with an integrated intensity of $4.7\pm
     0.9$~Jy~km~s$^{-1}$. For reference, the \textit{dashed line}
     shows the best-fitting Gaussian model for these data.}
  \label{fig:oistack}
\end{figure}

\noindent
For the weighting we take the measured rms of the spectra, $\sigma_i$,
and assume weighting, $w_i = 1/\sigma_i$ and $w_i = 1/\sigma_i^2$. Both weighting
schemes give similar results, and in Figure~\ref{fig:oistack} we plot
the results obtained assuming $1/\sigma_i^2$. 
The average infrared luminosity of the three galaxies is $(9.3\pm 2.1)\times 10^{11}$~L$_{\odot}$.  

The results of our stacking analysis reveal a significant detection of the
\hbox{[OI]}63\,$\mu$m line emission with an integrated intensity, $4.7
\pm 0.9$~Jy~km~s$^{-1}$. As all three galaxies are at a similar
redshift, we can assume a common luminosity distance and therefore
calculate a line luminosity from the integrated intensity of the
stacked spectrum,
L$_{\rm [OI]63\mu m} = (1.1 \pm 0.2)\times 10^9$~L$_{\odot}$. 

\subsection{Luminosity ratios and PDR modelling}

Assuming that the line emission is cospatial with the
  thermal dust continuum emission, 
we calculate the ratio of the [OI]63$\mu$m line luminosity in BzK-21000 to
its total 8-1000$\mu$m infrared luminosity measured by Magdis et al.\
(2012). These authors use photometry from \textit{Herschel} PACS and SPIRE to
constrain the full infrared spectral energy distributions of the targets in our
sample. The measured luminosities are presented in
Table~\ref{tab:obs}. From this we calculate a luminosity ratio, $L_{\rm [OI]63\,\mu
  m}/L_{\rm IR} = (1.8 \pm 0.3)\times 10^{-3}$, for BzK-21000. 

\begin{figure}
\includegraphics[width=1.1\columnwidth]{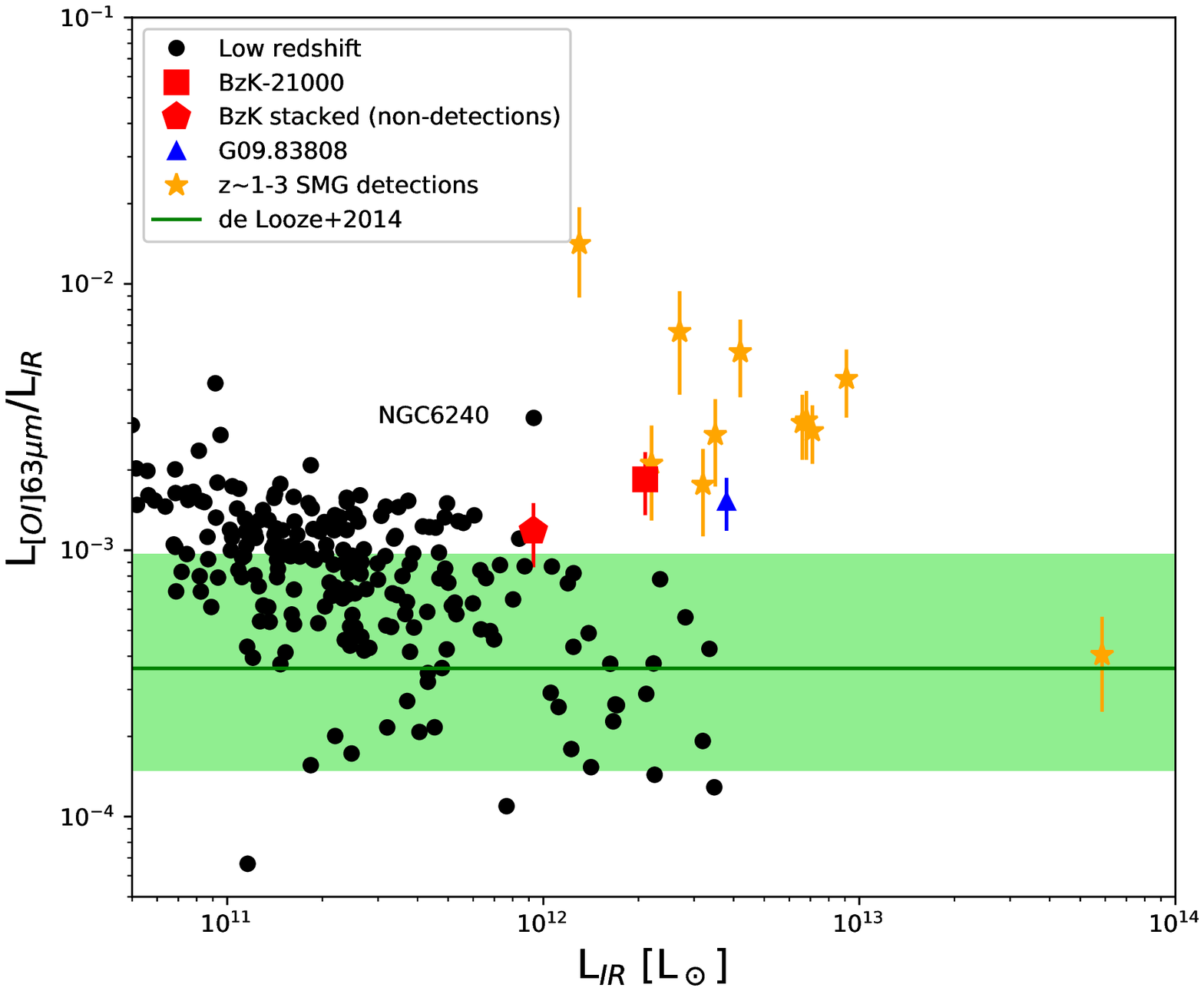}
   \caption{We plot the luminosity ratio, $L_{\rm [OI]63\,\mu m
     }/L_{\rm IR}$ versus L$_{\rm IR}$ for the GOALS sample of
     nearby galaxies (Diaz-Santos et al.\ 2017), compared to those galaxies which have been
    detected at high-redshift. The \textit{stars} show the luminosity
    ratios of lensed and unlensed submm galaxies at $z \sim 1-3$ (Coppin
    et al.\ 2012; Brisbin et al.\ 2015; Zhang et al.\ 2018), while Rybak et al.\ (2020)
    detect line emission in G09.83808 at $z \sim 6$ using
    the APEX telescope. Where applicable, the luminosities have been
    corrected for gravitational lensing. For reference we also plot
    the $[OI]63\,\mu$m-to-IR relationship for local galaxies (de
    Looze et al.\ 2014).
The error bars include an additional 20\% to account for instrumental calibration uncertainty 
on the line measurements. }
  \label{fig:loilfir}
\end{figure}

In Figure~\ref{fig:loilfir} we plot the [OI]63$\mu m$ line-to-infrared luminosity
ratio of galaxies as a function of infrared luminosity. The luminosity
ratio for BzK-21000 is compared to the low-redshift GOALS sample of star-forming galaxies
and AGN (Diaz-Santos et al.\ 2017), along with the average ratio derived
by de Looze et al.\ (2014). Also plotted are the luminosity ratios of the submm
 galaxies at $z \sim 1-3$ detected using \textit{Herschel}-PACS
(Coppin et al.\ 2012; Brisbin et al.\ 2015; Zhang et al.\ 2018), where three of the 8 to 1000~$\mu$m
infrared luminosities are from Swinbank et al.\ (2014), and the APEX telescope detection of
[OI]63$\mu$m emission in a gravitationally lensed, dusty galaxy at $z
\sim 6$ (Rybak et al.\ 2020). All of our error bars
include an additional 20\% to account for instrumental calibration uncertainty 
on the line measurements. 

The high-redshift galaxies detected in [OI]63$\mu m$ line emission
tend to exhibit a higher line-to-infrared luminosity ratio than that of typical
galaxies in the nearby Universe. One of the clear exceptions to this 
is NGC6240, exhibiting both a high infrared luminosity
($\sim9\times10^{11}$~L$\odot$) and strong [OI]63$\mu$m line
emission ($\sim 2.9\times 10^{9}$~L$_{\odot}$). This dual AGN is
known to have a very warm and dense ISM, as revealed by studies of
molecular CO line emission and dense gas tracers HCN and HCO$^+$
(e.g. Greve et al.\ 2009; Meijerink et al.\ 2013; Scoville et al.\ 2015; Treister et al.\ 2020) . It also appears to have
a nuclear outflow traced by molecular gas (van der Werf et al.\ 1993; Iono et al.\
2007; Feruglio et al.\ 2013; Cicone et al.\ 2018). It is possible
that the extreme physical conditions of the ISM of NGC6240 may
have some similarities with that of high-redshift submm starburst galaxies and AGN detected
in the [OI]63$\mu$m line, however we note that both high-redshift
submm galaxies and main-sequence galaxies generally exhibit lower CO
line excitation. The presence of an AGN may be a
contributing factor, as up to four of the $z \sim 1 - 3$ submm galaxies with
strong [OI]63$\mu$m line emission are thought to contain an
AGN. The optical and UV lines in SMMJ030227.73 +000653.5 suggest that
an AGN is present (Swinbank et al.\ 2004; Takata et al.\ 2006), while
SDSS~J120602.09+514229.5 shows weak evidence in the form of a strong
[S~IV] line and hot mid-infrared colours (Fadely et al.\
2010). LESS66 has a \textit{Chandra} X-ray counterpart (Wang et al.\
2013) and MIPS~22530 is tentatively believed to host an AGN based on 
an analysis of its radio emission (Sajina et al.\ 2008).

Another possible interpretation of the high [OI]63$\mu m$ line
luminosity observed in submm galaxies and BzK-21000, is that this
emission arises from an extended reservoir of cool and low density, neutral
gas within these galaxies. Such a scenario would be consistent with the low molecular
CO line excitation observed in main sequence galaxies (e.g. Daddi et al.\
2008, 2010a, 2015, Dannerbauer et al.\ 2009; Aravena et al.\ 2010), and the extended reservoirs of
cold molecular gas traced by CO~\textit{J}=1-0 in some submm
galaxies (e.g. Ivison et al.\ 2010, 2011; Riechers et al.\ 2011). Blind mm and
cm-wavelength surveys of CO line emission in high-redshift galaxies
have found that galaxies selected via CO~\textit{J}=1-0 line emission
have a lower excitation, on average, than those selected through the
CO~\textit{J}=3-2 line (Riechers et al.\ 2020). These low excitation
galaxies may also be strong [OI]63$\mu m$ line emitters, with some
fraction of the neutral atomic gas arising from clumps of denser gas.

The results presented here are similar to what has been found by
  previous studies of far-infrared line emission in star-forming galaxies, where the
  line-to-infrared luminosity ratio shows a deficit that
  shifts to higher luminosities with redshift (e.g. Grazi\'a-Carpio et
  al.\ 2011). Such a redshift trend can be removed by plotting the 
  dependance of the luminosity ratio against the $\rm L_{FIR}/M_{H_2}$ ratio (related 
  to the star-formation efficiency). This can be understood by
  considering that the majority of
  galaxies studied in far-infrared line emission at high-redshift 
  are starburst galaxies which exhibit a high star-formation
  efficiency compared to main sequence galaxies. Observing the [CII] line emission in
  main-sequence galaxies over a range in luminosities and redshifts,
  Zanella et al.\ (2018) showed that the luminosity in this line is strongly correlated 
  with total molecular gas. As such, the $\rm L_{[CII]}/ L_{FIR}$
  ratio could be interpreted as the gas-depletion
  timescale for galaxies such as those in our sample.

\begin{figure}
\includegraphics[width=1.\columnwidth]{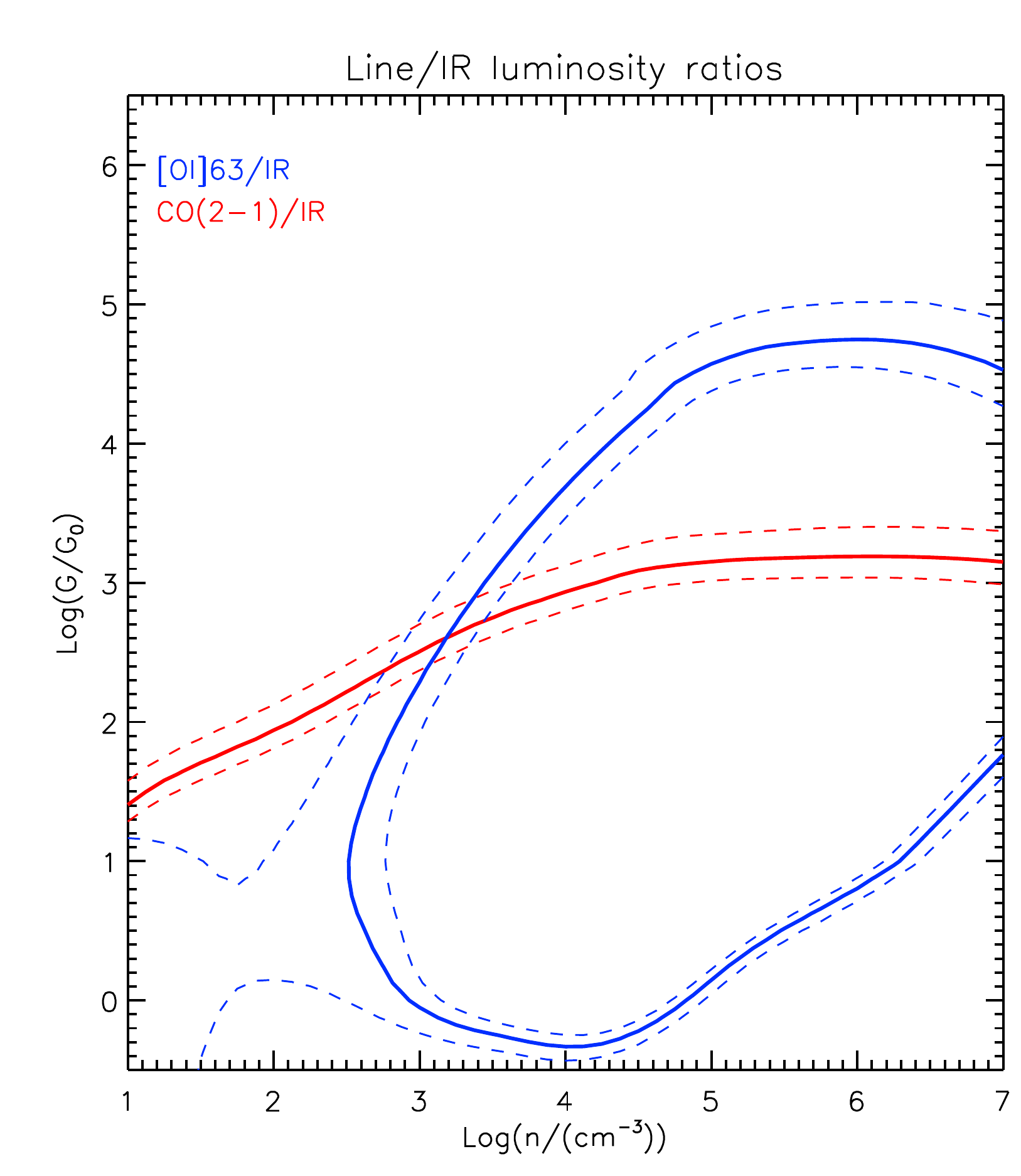}
   \caption{PDR model parameter space constrained by our measured
     luminosity ratios for BzK-21000. The CO~\textit{J}=2-1 line
     luminosity is from Daddi et al.\ (2010a). Using the PDR toolbox
     models (Pound \& Wolfire 2008 and Kaufman et al.\ 2006) 
     updated by the authors (Wolfire priv. communication), the UV radiation field within the
     [OI]63$\mu$m emission line region is
     estimated to be, $G \sim 320 G_0$, at a gas density of, $n \sim
     1800$~cm$^{-3}$, if concomittant with the low-\textit{J} CO line emission.}
  \label{fig:PDRmodels}
\end{figure}

Finally, we consider the constraints that our luminosity ratios imply
for the physical conditions within the ISM of BzK-21000 where the
[OI]63$\mu$m emission arises. We assume that the [OI]63$\mu$m line
  and thermal dust continuum emission regions are cospatial
with the CO~\textit{J}=2-1 line emission measured by Daddi et al.\
(2010a), acknowledging that high angular resolution imaging would be
required to verify this assumption. To investigate the PDRs we adopt 
updated PDR models based on the online PDR toolbox (Pound \& Wolfire
2008 and Kaufman et al. 2006, Wolfire priv. communication). 
These models have recently been updated to reflect the chemistry 
and reaction rates noted in Hollenbach et al. (2012), and Neufeld \&
Wolfire (2016), as well as photodissociation and ionization rates from
Heays et al. (2017), and the collisional excitation of OI from Lique
et al. (2018). These models assume a simple slab PDR geometry
 illuminated on one side. For an ensemble of PDR cloudlets externally 
illuminated, optically thin emission will be observed from both sides, 
while optically thick emission is only observed from the front side. 
Therefore, as recommended in Kaufman et al. (1999) and commonly
practiced in PDR analysis (e.g. Hailey-Dunsheath et al 2010; Stacey et
al. 2010), we have doubled the observed line fluxes of [OI] and
CO when fitting the data to account for their expected optical
thickness. Based on these models, the L$_{\rm [OI]63\mu m}/$L$_{\rm IR}$ and
L$_{\rm CO}/$L$_{\rm IR}$ luminosity ratios suggest a UV radiation
field, $G \sim 320 G_0$, and gas density, $n \sim 1800$~cm$^{-3}$
(Figure~\ref{fig:PDRmodels}). 
This gas density is broadly consistent with the Large Velocity Gradient
(LVG) models fit to the observed CO~\textit{J}=2-1 and \textit{J}=3-2
line intensities (Dannerbauer et al.\ 2009; Daddi et al.\ 2015). Although these gas
densities are low compared to what is typically inferred 
 for submm luminous galaxies observed in [CII] and
[OI]63$\mu$m (Brisbin et al.\ 2015), the radiation field strengths are
similar. Further constraints on the ISM conditions within BzK-21000 would be possible with detections 
of other FIR lines like [CII] or [NII]. 

\section{Conclusions}

We present \textit{Herschel}-PACS spectroscopy of four BzK-selected star-forming galaxies at $z \sim 1.5$. One of our targets, BzK-21000
at $z = 1.5213$ is detected with an [OI]63$\mu$m line luminosity,
L$_{\rm [OI]63\,\mu m} = (3.9\pm 0.7)\times 10^9$~L$_{\odot}$. A spectral stacking analysis of the data from the three non-detections
reveals a significant signal, implying  L$_{\rm [OI]63\,\mu m} =
(1.1\pm 0.2)\times10^9$~L$_{\odot}$. 

The line-to-total infrared luminosity ratio in BzK-21000 is similar to that of a
dusty, $z\sim 6$ galaxy (Rybak et al.\ 2020), but lower than that
typically observed in massive submm galaxies at $z \sim 1 -
3$. Combined with PDR models, the relative strengths of the
[OI]63$\mu$m and CO~\textit{J}=2-1 lines compared to the infrared
luminosity imply a UV field intensity, $G\sim 320G_0$, and a
gas density, $n \sim 1800~$cm$^{-3}$. The gas density is low compared
to the average determined for 
more massive submm galaxies observed in [OI]63$\mu$m at high-redshift
(Brisbin et al.\ 2015). 

Given the observed intensity of the [OI]63$\mu$m line emission in the
BzK-selected star-forming galaxies studied here, it is likely that
ALMA would be a powerful instrument for studying this line in more
distant, main sequence galaxies. Beyond redshifts, $z \ga 4$, this line
is redshifted into the ALMA band 10 receiver range. The $602 -
720$~GHz band 9 receivers can observe this line in galaxies
during the Epoch of Reionization at $z \ga 6$,  and have been
 used to study the [CII]158$\mu$m line in lower redshift galaxies
 (e.g. Schaerer et al.\ 2015; Zanella et al.\ 2018; Lamarche et al.\ 2018).

\section*{Acknowledgements}

We thank the anonymous referee for a thorough review of the original
manuscript, and for useful feedback. In addition, we thank
 Kristen Coppin and Mark Swinbank for helpful
discussions. MA and this work have been supported by grants
``CONICYT+PCI+REDES 19019"  and ``CONICYT + PCI + INSTITUTO MAX PLANCK
DE ASTRONOMIA MPG190030''. D.R. acknowledges support from the National Science Foundation
   under grant numbers AST-1614213 and AST-1910107. D.R. also
   acknowledges support from the Alexander von Humboldt Foundation
   through a Humboldt Research Fellowship for Experienced Researchers.
PACS has been developed by a consortium of institutes led
by MPE (Germany) and including UVIE (Austria); KU Leuven, CSL, IMEC
(Belgium); CEA, LAM (France); MPIA (Germany); INAF-IFSI/OAA/OAP/OAT,
LENS, SISSA (Italy); IAC (Spain). This development has been supported
by the funding agencies BMVIT (Austria), ESA-PRODEX (Belgium),
CEA/CNES (France), DLR (Germany), ASI/INAF (Italy), and CICYT/MCYT
(Spain).

\section*{Data Availability}

The data underlying this article can be accessed from the ESA
\textit{Herschel} science archive: \textit{http://archives.esac.esa.int/hsa/whsa/}. The derived data generated in this research will be shared on reasonable request to the corresponding author.

%%%%%%%%%%%%%%%%%%%%%%%%%%%%%%%%%%%%%%%%%%%%%%%%%%



%\bibliographystyle{mnras}


%%%%%%%%%%%%%%%%%%%%%%%%%%%%%%%%%%%%%%%%%%%%%%%%%%

%%%%%%%%%%%%%%%%% APPENDICES %%%%%%%%%%%%%%%%%%%%%

%\appendix



%%%%%%%%%%%%%%%%%%%%%%%%%%%%%%%%%%%%%%%%%%%%%%%%%%

% Don't change these lines
\bsp	% typesetting comment
\label{lastpage}
\end{document}